\documentclass[aps,prl,amsmath,amssymb,floatfix,twocolumn,amsmath,superscriptaddress,twocolumn,nofootinbib,tighten,letterpaper]{revtex4-2}
\usepackage[colorlinks,linkcolor=blue,citecolor=blue,urlcolor=blue]{hyperref}
\usepackage{multirow}
\usepackage{subfigure}
\usepackage{color}
\usepackage{mathrsfs}
\usepackage{hyperref}
\usepackage[normalem]{ulem}
\usepackage{bm}

\usepackage{amssymb}
\usepackage{amsmath}
\renewcommand\vec[1]{\ensuremath\boldsymbol{#1}}

\usepackage{tabularray}

\usepackage{array} 
\newcolumntype{P}[1]{>{\centering\arraybackslash}p{#1}}

\definecolor{RowColor}{rgb}{0.88,1,0.9}

\usepackage{amsfonts, relsize, color, mathtools, physics}
\usepackage{graphics}
\usepackage{graphicx}
\usepackage{subfigure}
\usepackage{color}
\usepackage{comment}

\begin{document}

\title{Hidden-ordered Dirac fermions}

\author{Bitan Roy}
\affiliation{Department of Physics, Lehigh University, Bethlehem, Pennsylvania, 18015, USA}

\date{\today}

\begin{abstract}
I propose a Hermitian extension of the Lorentz-symmetric Dirac theory by complementing the associated Hamiltonian with another \emph{masslike} anticommuting Dirac operator. The resulting theory manifests the iconic linear energy-momentum relationship in any dimension ($d$) and hence the emergent nodal quasiparticle excitations are named \emph{hidden-ordered Dirac fermions}, which are symmetry protected and their responses are analogous to those in original Dirac systems, however, in terms of a renormalized (due to the hidden ordering) Fermi velocity. Typically, such a hidden ordering pushes any quantum phase transition into an insulation toward even stronger coupling in any $d>1$. However, depending on the internal algebra between the candidate insulating order parameter and masslike Dirac operator, the hidden-ordering may survive or disappear near the corresponding itinerant quantum critical point. I construct lattice models for such hidden-ordered massless Dirac fermions and outline promising platforms (numerical and experimental) to test these predictions.      
\end{abstract}

\maketitle

\emph{Introduction}.~Linear energy-momentum relationship is critical to ensure the Lorentz symmetry in the Dirac theory~\cite{Dirac:1, Dirac:2}. Furthermore, the discrete charge conjugation (${\mathcal C}$), parity or inversion (${\mathcal P}$), and time-reversal (${\mathcal T}$) symmetries restrict the associated Dirac Hamiltonian for relativistic fermions to its canonical form~\cite{peskin}. Specifically, the Dirac theory permits two quantities that transform as scalars under the Lorentz transformation; (a) the Dirac Hamiltonian that is linear in momentum and (b) a constant mass terms. Both of them are accompanied by appropriate $\gamma$ matrices, which mutually anticommute. In terms of such Lorentz-Dirac scalars, here I propose a Lorentz-symmetric Hermitian extension of the Dirac Hamiltonian for massless quasi-relativistic excitations that continues to display the hallmark linear energy-momentum relation in any dimension ($d$).

\emph{Construction}.~The Dirac Hamiltonian for quasi-relativistic fermions with the Fermi velocity $v_{_{\rm F}}$ takes the universal form $H_{\rm Dir} (\vec{k}) = v_{_{\rm F}} \left( \Gamma_j k_j \right)$, where $j=1,\cdots d$ and a summation over repeated indices is assumed. The Hermitian $\Gamma$ matrices satisfy the anticommuting Clifford algebra $\{ \Gamma_j, \Gamma_k \}= 2 \delta_{jk}$, where $\delta_{jk}$ is the Kronecker delta symbol, yielding the energy spectra $\pm E_{{\vec k}}$ manifesting the linear energy-momentum relation $E_{{\vec k}}= v_{_{\rm F}} |\vec{k}|$. The dimensionality and explicit representation of the $\Gamma$ matrices depend on $d$ and microscopic details, about which more in a moment.  In this language, any Dirac mass is represented by a Hermitian matrix $M$, satisfying $\{M, \Gamma_j \}=0$ for any $j$ and $M^2=\Gamma_0$ (identity matrix). In terms of these two ingredients, I propose the following Lorentz invariant extended Dirac operator
\begin{equation}~\label{eq:HOHamiltonian}
H_{\rm HO} (\vec{k})=H_{\rm Dir} (\vec{k}) + i \alpha M H_{\rm Dir} (\vec{k}),
\end{equation}
where $\alpha$ is a dimensionless quantity. The energy spectra of $H_{\rm HO} (\vec{k})$ are $\pm E_{\rm HO}(\vec{k})$ with $E_{\rm HO}(\vec{k})= v_{_{\rm F}} \sqrt{1+\alpha^2} |\vec{k}|$. Therefore, $H_{\rm HO} (\vec{k})$ continues to display the characteristic linear energy-momentum relation in terms of a renormalized Fermi velocity $v^{\rm R}_{_{\rm F}}= v_{_{\rm F}} \sqrt{1+ \alpha^2}$, despite containing a mass matrix ($M$) therein which typically stems from ordering, devoid of some discrete and/or continuous symmetries. So, I name such nodal quasi-relativistic excitations \emph{hidden-ordered Dirac fermions} (HODF) and showcase the outcomes, summarized below. In monolayer graphene (a planar Dirac system), it is conceivable to find gapless superconducting phases with pairing amplitude $\alpha$ for which the effective single-particle Bogoliubov de-Gennes Hamiltonian takes the form of $H_{\rm HO} (\vec{k})$~\cite{Roy-Herbut-2010:PRB, Uchoa-CastroNeto-2007:PRL}.

\emph{Key results}.~The Hamiltonian for HODF is unitarily equivalent to a \emph{scaled} Dirac Hamiltonian according to 
\begin{equation}
H_{\rm HO} (\vec{k})= e^{i M \theta/2} \: \left\{ \sqrt{1+\alpha^2} \; H_{\rm Dir} (\vec{k}) \right\} \: e^{-i M \theta/2},
\end{equation}
where $\theta=\tan^{-1}(\alpha)$. The gapless nature of such HODF is shown to be symmetry protected, despite $H_{\rm HO} (\vec{k})$ containing a symmetry-breaking mass matrix ($M$). Thermodynamic, transport, and elastic responses of massless HODF show the same power-law scaling as the original Dirac systems, however, in terms of $v^{\rm R}_{_{\rm F}}$.

The enhanced Fermi velocity yielding a larger band width defers any spontaneous symmetry breaking in hidden-ordered Dirac systems toward a stronger coupling in comparison to that in conventional Dirac systems. Here, this outcome is shown by considering only mass orders that are represented by matrices ($N$) that fully anticommute with $H_{\rm Dir}(\vec{k})$. However, the quantum critical phenomena in the close proximity to mass orderings crucially depend on the internal algebra between the candidate mass order matrix $N$ and the mass matrix appearing in $H_{\rm HO} (\vec{k})$, namely $M$. When $[N,M]=0$, thus the ordered phase earns the name commuting class mass (CCM), the parameter associated with the hidden-ordering ($\alpha$) remains \emph{marginal}. If, on the other hand, $\{ N, M \}=0$, for which the ordered phase is coined anticommuting class mass (ACM), the parameter $\alpha$ becomes marginally \emph{irrelevant} near such ordering. Alternatively, the hidden-ordering disappears near the ACM ordering. Next these results are discussed in detail.

\emph{Symmetry protection}.~In $d=2$ the minimal representation of pseudo-spin Dirac fermions in any lattice-based spinless system (such as graphene) is four and the $\Gamma$ matrices appearing in $H_{\rm Dir}(\vec{k})$ are four-dimensional. Without loss of generality, I choose $\Gamma_1$ and $\Gamma_2$ to be purely imaginary~\cite{okubo}, for which ${\mathcal T}= {\mathcal K}$, where ${\mathcal K}$ is the complex conjugation, such that ${\mathcal T}^2=+1$, as it should be for spinless fermions. The (antiunitary) particle-hole symmetry is generated by ${\mathcal C}={\boldsymbol \Gamma} {\mathcal K}$, where ${\boldsymbol \Gamma} \in \{ \Gamma_3, \Gamma_4, \Gamma_5, \Gamma_{12}\}$ with $\Gamma_{jk}=i \Gamma_j \Gamma_k$. Note that ${\boldsymbol \Gamma}$ is the set of matrices that fully anticommute with $H_{\rm Dir}(\vec{k})$ and thus represents all the mass orders. I choose $\Gamma_3$, $\Gamma_4$, and $\Gamma_5$ to be purely real and $\{ \Gamma_j \}$ with $j=1,\cdots, 5$ constitute the set of maximal five mutually anticommuting four-dimensional Hermitian matrices. The unitary particle-hole or sublattice symmetry is generated by ${\mathcal S}={\boldsymbol \Gamma}$. The generators of rotational symmetry about the $z$ axis is $\Gamma_{12}$. In addition, the two-dimensional Dirac Hamiltonian also enjoys a SU(2) chiral symmetry, generated by $\{\Gamma_{34}, \Gamma_{45}, \Gamma_{53} \}$. Hence, the triplet mass $\{ \Gamma_3, \Gamma_4, \Gamma_5 \}$ breaks (preserves) the chiral (time-reversal) symmetry, while the signlet mass $\Gamma_{12}$ preserves (breaks) the chiral (time-reversal) symmetry~\cite{graphenemicroscopic:1, graphenemicroscopic:2}. So, it is sufficient to choose $M=\Gamma_3$ and $\Gamma_{12}$ in Eq.~\eqref{eq:HOHamiltonian} to establish symmetry protection of nodal HODF in $d=2$.

With the former choice of $M$, $H_{\rm HO}(\vec{k})$ lacks the ${\mathcal T}$ symmetry, but possesses particle-hole symmetries, generated by ${\mathcal C}= \Gamma_4 {\mathcal K}, \Gamma_5 {\mathcal K}$ and ${\mathcal S}= \Gamma_3, \Gamma_{12}$, besides the rotational symmetry, and a reduced U(1) chiral symmetry under $\Gamma_{45}$. As shown in the Supplemental Material (SM), residual symmetries of $H_{\rm HO}(\vec{k})$ are sufficient to forbid the nucleation of any fermion bilinear~\cite{SM}. By contrast, when $M=\Gamma_{12}$, $H_{\rm HO}(\vec{k})$ possesses identical symmetries as $H_{\rm Dir}(\vec{k})$. Hence, formation of finite vacuum expectation value of any fermionic bilinear must be associated with the breakdown of some symmetry of the noninteracting system. Therefore, a collection of HODF in two spatial dimensions is always symmetry protected.

In $d=3$, I choose three four-dimensional $\Gamma$ matrices ($\Gamma_1$, $\Gamma_2$, and $\Gamma_3$) appearing in $H_{\rm Dir}(\vec{k})$ to be purely real. The time-reversal operator is then ${\mathcal T}= \Gamma_4 {\mathcal K}$ or $\Gamma_5 {\mathcal K}$. For either choice ${\mathcal T}^2=-1$ as it should be, since a four-component Dirac theory in $d=3$ involves relativistic spin-1/2 fermions with strong spin-orbit coupling~\cite{Dirac:1, Dirac:2, peskin}. The particle-hole symmetries are generated by ${\mathcal C}= \Gamma_{45} {\mathcal K}$ or ${\mathcal K}$ and ${\mathcal S}=\Gamma_4$ or $\Gamma_5$. The parity or inversion, under which $\vec{r} \to -\vec{r}$ and $\vec{k} \to -\vec{k}$, is generated by ${\mathcal P}=\Gamma_4$ or $\Gamma_5$. The three-dimensional Dirac Hamiltonian also enjoys O(3) rotational symmetry generated by $\{ \Gamma_{12}, \Gamma_{23}, \Gamma_{31} \}$ and a U(1) chiral symmetry under $\Gamma_{45}$.

Two mutually anticommuting Hermitian mass matrices $\Gamma_4$ and $\Gamma_5$ are then purely imaginary and they break the U(1) chiral symmetry. Thus in the Hamiltonian construction of HODF I exclusively choose $M=\Gamma_4$ to scrutinize its symmetry protection, for which $H_{\rm HO}(\vec{k})$ enjoys the ${\mathcal T}$ symmetry generated by $\Gamma_4 {\mathcal K}$, ${\mathcal S}$ symmetry under $\Gamma_4$, ${\mathcal C}$ symmetry under ${\mathcal K}$, and the ${\mathcal P}$ symmetry generated by $\Gamma_4$, besides the spatial rotational O(3) symmetry. Such symmetries are sufficient to disallow the nucleation of any fermion bilinear unless some symmetry of the system is compromised~\cite{SM}. Therefore, a collection of three-dimensional HODF is also symmetry protected. 

\emph{Responses}.~Due to the linear energy-momentum relationship, yielding the dynamic scaling exponent $z=1$, the density of states (DoS) for HODF scales as $\rho(E)\sim |E|^{d-1}/[v^{\rm R}_{_{\rm F}}]^d$, which vanishes for any $d>1$. Thus the increased Fermi velocity for HODF causes a parametric suppression of the DoS near zero-energy without altering its power-law behavior, which also reflects in the scaling of the specific heat $C_V \sim \left( T/v^{\rm R}_{_{\rm F}} \right)^d$.

The response of HODF to external electromagnetic field can be captured in terms of the optical conductivity at finite frequency ($\omega$) in $d$ dimensions $\sigma^{(d)}(\omega)$, which can be computed using the Kubo formula, yielding~\cite{SM} 
\begin{equation}
\sigma^{(2)}(\omega)= N_f \frac{e^2}{h} \frac{\pi}{4} 
\:\: \text{and} \:\:
\sigma^{(3)}(\omega)= N_f \frac{e^2}{h} \frac{\omega}{6 v^{\rm R}_{_{\rm F}}}
\end{equation}
respectively, in two and three dimensions, where $N_f$ is the number of four-component fermion flavors. Therefore, the optical conductivity for HODF gets suppressed due to the enhanced Fermi velocity in comparison to that for its conventional counterpart in $d=3$~\cite{OC:1, OC:2}. By contrast, in $d=2$ hidden-ordered and conventional Dirac fermions share the same value of the optical conductivity~\cite{OC:3}.

Finally, I note that the frequency-dependent optical shear viscosity for noninteracting HODF is captured by a single independent component due to the underlying rotational symmetry of the system, and in $d$ spatial dimensions it is given by $\eta^{(d)}_{jklm} (\omega)= {\mathcal P}_{jklm} \; \eta^{(d)}(\omega)$, where ${\mathcal P}_{jklm}=\delta_{jl} \delta_{km} + \delta_{jm} \delta_{kl} - (2/d) \delta_{jk} \delta_{lm}$, 
\begin{equation}
\eta^{(2)}(\omega) = \frac{N_f}{128} \left( \frac{\omega}{v^{\rm R}_{_{\rm F}}}\right)^2,
\; \text{and} \;
\eta^{(3)}(\omega) = \frac{N_f}{320 \pi} \left( \frac{\omega}{v^{\rm R}_{_{\rm F}}}\right)^3.
\end{equation}
Therefore, optical shear viscosity for hidden-ordered and conventional Dirac fermions follows the same scaling behavior, but in terms of $v^{\rm R}_{_{\rm F}}$ and $v_{_{\rm F}}$, respectively~\cite{SM, viscosity:1, viscosity:2}.

\emph{Mean-field susceptibility}.~Although due to the vanishing DoS, symmetry-protected HODF are stable against sufficiently weak interactions in any $d>1$, strong enough Hubbardlike interactions can trigger quantum phase transitions (QPTs) toward the formation of various insulating phases in the system. In traditional Dirac systems, such ordered states are described by Hermitian matrices ($N$) that fully anticommute with $H_{\rm Dir}(\vec{k})$, and are named mass order. Here I exclusively focus on the mass orders as they become energetically most favored at and near zero temperature by pushing all the filled states at negative energies down and opening up an isotropic gap near the Dirac points at half filling. In conventional Dirac systems, all the mass orders receive equal propensity due to their \emph{universal} anticommutation relation $\{ N, H_{\rm Dir}(\vec{k}) \}=0$, irrespective of the choice on $N$.

However, appearance of a mass matrix in $H_{\rm HO}(\vec{k})$, hidden ordering causes a fragmentation within the set of mass matrices into the families of CCM and ACM. Propensity toward the nucleation of either class of mass orderings can be assessed from the corresponding bare mean-field susceptibility at zero external frequency and momentum ($\chi$). The critical coupling for any ordering is $\tilde{g}_\star \propto \chi^{-1}$, and for CCM and ACM orderings~\cite{SM} 
\begin{equation}
\frac{\tilde{g}^{\rm CCM}_\star (\alpha)}{\tilde{g}^{\rm CCM}_\star (0)}=\left( 1 + \alpha^2 \right)^{1/2}
\: \text{and} \:\:
\frac{\tilde{g}^{\rm ACM}_\star (\alpha)}{\tilde{g}^{\rm ACM}_\star (0)}=\left( 1 + \alpha^2 \right)^{3/2},
\end{equation}
respectively. Therefore, both types of mass orderings get pushed toward stronger coupling due to the hidden ordering in the systems (finite $\alpha$), a phenomenon coined here as `\emph{inverse catalysis of spontaneous symmetry breaking}'. But, the critical coupling for the ACM ordering increases more rapidly in comparison to that for the CCM ordering. Such a concrete prediction from a mean-field analysis can be tested from lattice based numerical simulations within the Hartree or Fock approximation.

\emph{Yukawa theory}.~Next I proceed to expose the quantum critical phenomena near CCM and ACM orderings, at first focusing on the fate of the hidden ordering (captured by $\alpha$) near the associated quantum critical points (QCPs). The effective field theory describing such QPTs takes the form of the Gross-Neveu-Yukawa (GNY) model with the corresponding imaginary time ($\tau$) Euclidean action assuming the form $S=\int d\tau \int d^d \vec{x} \left[ {\mathcal L}_{\rm F}+ {\mathcal L}_{\rm BF}+ {\mathcal L}_{\rm B} \right]$. The fermionic Lagrangian density reads as 
\begin{equation}
{\mathcal L}_{\rm F}= \Psi^\dagger(\tau,\vec{x}) \left[ \partial_\tau + H_{\rm HO} (\vec{k} \to -i {\boldsymbol \nabla})\right] \Psi (\tau,\vec{x}),
\end{equation}
while the coupling between massless HODF with the order parameter bosonic field (${\boldsymbol \Phi}$) through the Yukawa interaction ($g$) is captured by the Lagrangian density
\begin{equation}
{\mathcal L}_{\rm BF}= g \:  \sum^{N_b}_{\kappa=1} \Phi_\kappa(\tau,\vec{x}) \Psi^\dagger(\tau,\vec{x}) N_\kappa \Psi (\tau,\vec{x}),
\end{equation}
where the mass order is assumed to have $N_b$ components. The dynamics of the order parameter field is described by the $\Phi^4$ theory, with the Lagrangian density
\begin{equation}
{\mathcal L}_{\rm B} = \sum^{N_b}_{\kappa=1} \bigg[\frac{\Phi_\kappa}{2} \left(m^2_{_{\rm B}} - \partial^2_\tau - v^2_{_{\rm B}} \sum^{d}_{j=1} \partial^2_j \right) \Phi_\kappa
+ \frac{\lambda}{4!} \left[ \Phi^2_\kappa \right]^2 \bigg],
\end{equation}
where $\Phi_\kappa \equiv \Phi_\kappa(\tau,\vec{x})$, $v_{_{\rm B}}$ is the velocity of the bosonic order parameter fields with $v_{_{\rm B}} \neq v_{_{\rm F}}$ (typically), $\lambda$ is the four boson interaction coupling, and $m^2_{_{\rm B}}$ is the bosonic mass that serves as the tuning parameter for the QPT with the GNY-QCP located on the $m^2_{_{\rm B}}=0$ hyperplane. Both Yukawa ($g$) and four-boson ($\lambda$) couplings are dimensionless in $d=3$, thereby facilitating a controlled $\epsilon$ expansion about the upper critical three spatial dimensions to capture the emergent quantum critical phenomena near the CCM and ACM ordering with $\epsilon=3-d$~\cite{justin}.

\begin{figure}[t!]
\includegraphics[width=1.00\linewidth]{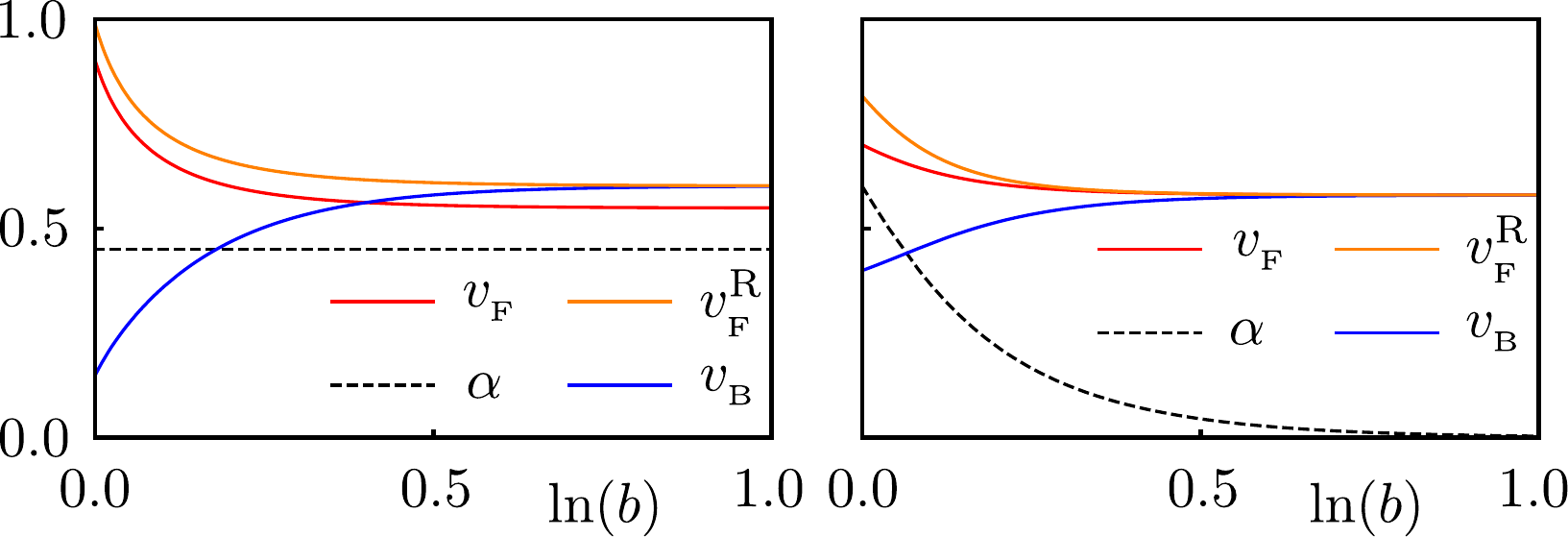}
\caption{Renormalization group flows of Fermi velocities ($v_{_{\rm F}}$ and $v^{\rm R}_{_{\rm F}}$), bosonic velocity ($v_{_{\rm B}}$), and $\alpha$ quantifying the hidden-ordering in the critical hyperplane ($m^2_{_{\rm B}}=0$) with Yukawa coupling $g^2=1$, number of four-component fermion flavors $N_f=1$, and bosonic order-parameter component $N_b=1$ near (a) the commuting class mass ordering with the bare values (denoted by the superscript `0') of $v^0_{_{\rm F}}=0.90$, $\alpha^0=0.45$, and $v^0_{_{\rm B}}=0.15$ and (b) the anticommuting class mass ordering with the bare values of $v^0_{_{\rm F}}=0.70$, $\alpha^0=0.60$, and $v^0_{_{\rm B}}=0.40$. Here, $b$ is the renormalization group time.           
}~\label{fig:RGFlows}
\end{figure}

At first, I consider the renormalization group (RG) flow of the velocities ($v_{_{\rm F}}$ and $v_{_{\rm B}}$) and $\alpha$ (hidden ordering). To this end, the fermionic and bosonic self-energy diagrams are computed, leading to the RG flow equations 
\begin{equation}
\beta_{v_{_{\rm F}}}= \frac{4}{3} \frac{ N_b \; g^2 \; v_{_{\rm F}}}{v_{_{\rm B}} \left[ v_{_{\rm F}} \sqrt{1+\alpha^2} + v^2_{_{\rm B}} \right]^2} \left( \frac{v_{_{\rm B}}}{v_{_{\rm F}} \sqrt{1+\alpha^2}}-1 \right)
\end{equation}
near both types of mass orderings, and
\begin{eqnarray}
\beta_\alpha = 0, \:
\beta_{v_{_{\rm B}}} = \frac{N_f \; g^2 \; v_{_{\rm B}}}{2 v^3_{_{\rm F}}(1+\alpha^2)^{3/2}} 
\left( \frac{v^2_{_{\rm F}}(1+\alpha^2)}{v^2_{_{\rm B}}} -1 \right) 
\end{eqnarray}
near the CCM ordering and 
\begin{eqnarray}
\beta_\alpha &=& - \frac{4}{3} \frac{N_b \; g^2 \; \alpha}{v_{_{\rm B}} \left[ v_{_{\rm F}} \sqrt{1+\alpha^2} + v^2_{_{\rm B}} \right]^2} \left( 1 + \frac{2 v_{_{\rm B}}}{v_{_{\rm F}} \sqrt{1+\alpha^2}} \right), \nonumber \\
\beta_{v_{_{\rm B}}} &=& \frac{N_f \; g^2 \; v_{_{\rm B}}}{2 v^3_{_{\rm F}}(1+\alpha^2)^{3/2}} 
\left( \frac{1}{1+\alpha^2} - \frac{v^2_{_{\rm F}}(3-2\alpha^2)}{v^2_{_{\rm B}}}\right) 
\end{eqnarray}
near ACM ordering in terms of the dimensionless Yukawa coupling constant $g^2/(8 \pi^2) \to g^2$~\cite{SM}. The numerical solutions of these coupled RG flow equations are shown in Fig.~\ref{fig:RGFlows} from which I arrive at the following conclusions.

\emph{Emergent Lorentz symmetry}.~Near the CCM ordering, $\alpha$ remains \emph{marginal} and under coarse grain the renormalized Fermi velocity $v^{\rm R}_{_{\rm F}}$ and the bosonic velocity $v_{_{\rm B}}$ approach a common terminal value, as shown in Fig.~\ref{fig:RGFlows}(a) by assuming that the bare Fermi velocity $v_{_{\rm F}}$ (and thus $v^{\rm R}_{_{\rm F}}$) is larger than the bare value of $v_{_{\rm B}}$. I arrive at the same conclusion when $v_{_{\rm B}}>v_{_{\rm F}}$ and $v^{\rm R}_{_{\rm F}}$ at the bare level (not shown explicitly). Therefore, near the GNY-QCP associated with the CCM ordering, a hidden-ordered Dirac system features a \emph{unique} terminal velocity for all the participating degrees of freedom, thereby manifesting an emergent Lorentz symmetry.

Near the ACM ordering, on the other hand, $\alpha$ is marginally \emph{irrelevant} and under coarse grain $\alpha \to 0$. Concomitantly, $v^{\rm R}_{_{\rm F}} \to v_{_{\rm F}}$ as the system approaches the deep infrared regime and at the same time $v_{_{\rm F}} \to v_{_{\rm B}}$. I show such findings for $v^{\rm R}_{_{\rm F}} > v_{_{\rm B}}$ at the bare level in Fig.~\ref{fig:RGFlows}(b), which remain qualitatively unchanged when $v^{\rm R}_{_{\rm F}} < v_{_{\rm B}}$ (at the bare level) that is not shown explicitly. Therefore, in the close proximity to the ACM ordering a Lorentz symmetry emerges, manifesting in terms of a common terminal velocity for all the involved degrees of freedom. However, near the associated GNY-QCP, hidden ordering \emph{disappears} from the system.

\emph{Quantum criticality}.~With the emergence of the Lorentz symmetry being established I now focus on the Yukawa-Lorentz symmetric hyperplane, described by $v^{\rm R}_{_{\rm F}}=v_{_{\rm B}}$ near the CCM ordering and $v_{_{\rm F}}=v_{_{\rm B}}$ near the ACM ordering, to study the emergent quantum critical phenomena therein, which can be extracted from the following RG flow equations~\cite{SM} 
\begin{eqnarray}
\beta_{g^2} &=& \epsilon g^2 - (2N_f +4 - N_b) g^4 \nonumber \\ 
\text{and} \; \beta_{\lambda} &=& \epsilon \lambda - 4 N_f g^2 (\lambda-6 g^2) - \frac{8+N_b}{6} \lambda^2,
\end{eqnarray}
in terms of the dimensionless coupling constants $g^2/(8 \pi^2 v) \to g^2$ and $\lambda/(8 \pi^2 v) \to \lambda$ with $v=v^{\rm R}_{_{\rm F}}$ ($v=v_{_{\rm F}}$) in the CCM (ACM) ordering. Together the above two RG flow equations supports a QCP located at 
\begin{equation}
(g^2_\star, \lambda_\star)= \left(\frac{1}{2 N_f +4- n}, \frac{3\left[ b_2 + \sqrt{b^2_2+16 N_f b^2_1} \right]}{b_1 \; (2 N_f +4- N_b)} \right) \epsilon
\end{equation}
with $b_1=N_b+8$ and $b_2=4-2 N_f-N_b$, where both fermionic and bosonic fields become anomalous with the respective anomalous dimensions, given by $\eta_{_\Psi}=N_b g^2_\star/2$ and $\eta_{_\Phi}=2 N_f g^2_\star$. The residue of the fermionic quasiparticle pole vanishes near such a GNY-QCP as $Z \sim (m_{_{\rm F}})^{\eta_{_\Psi}/2}$, where $m_{_{\rm F}}$ is the fermionic mass that together with $m_{_{\rm B}}$ vanishes as $\left( m_{_{\rm F}}/m_{_{\rm B}} \right)^2 \sim g_\star/\lambda_\star$. Finally, the RG flow equation for the bosonic mass, given by 
\begin{equation}
\beta_{m^2_{_{\rm B}}}= m^2_{_{\rm B}} \left( 2- 2 N_f g^2 - \frac{2+n}{6} \lambda \right)
\end{equation}
in turn yields the correlation length exponent
\begin{equation}
\nu= \frac{1}{2} + \frac{N_f}{2} g^2_\star + \frac{2+n}{24} \lambda_\star.
\end{equation}
Therefore, in two spatial dimensions ($d=2$ or $\epsilon=1$), the QPT takes place through an interacting fixed point, where all the critical exponents acquire non-mean-field values, namely $\eta_{_\Psi} \sim \epsilon$, $\eta_{_\Phi} \sim \epsilon$, and $\nu-1/2 \sim \epsilon$. By contrast, in $d=3$ the QPT to any mass ordered phase occurs through the Gaussian fixed point, located at $(g^2_\star, \lambda_\star)=(0,0)$ and all the critical exponents receive mean-field values, namely $\eta_{_\Psi}=\eta_{_\Phi}=0$ and $\nu=1/2$.

\emph{Lattice model}.~I close the discussion with a concrete lattice model for HODF. For this purpose, consider the nearest-neighbor (NN) tight-binding Hamiltonian for spinless fermions on graphene's honeycomb lattice, a canonical setup that fosters massless Dirac fermions as emergent excitations near the corners of the hexagonal Brillouin zone. A two-component spinor is defined as $\Psi^\top_{\vec{q}}=(c^{\vec{q}}_A,c^{\vec{q}}_B)$, where $c^{\vec{q}}_A$ ($c^{\vec{q}}_B$) is the fermionic annihilation operator on the sites of the $A$ ($B$) sublattice of the underlying triangular Bravais lattice, $\vec{q}$ is the crystal momentum, and `$\top$' denotes transposition. In this basis, the Bloch Hamiltonian for the NN tight-binding model takes the form $H^{\rm latt}_{\rm Dir}(\vec{q})= t \{ \beta_1 \Re(f(\vec{q})) + \beta_2 \Im(f(\vec{q})) \}$, where $t$ is the NN hopping amplitude, $\{ \beta_\mu \}$ is the set of Pauli matrices operating on the sublattice index, $f(\vec{q})=\exp(i \vec{q} \cdot \vec{\delta}_1) + \exp(i \vec{q} \cdot \vec{\delta}_2) + \exp(i \vec{q} \cdot \vec{\delta}_3)$ with $\vec{\delta}_1=(1/\sqrt{3},1)a/2$, $\vec{\delta}_2=(1/\sqrt{3},-1)a/2$, and $\vec{\delta}_3=(-1/\sqrt{3},0)a$ as three NN vectors, and $a$ is the lattice spacing. In the same spinor basis, there exists a mass matrix $M_{\rm latt} \equiv \beta_3$ that corresponds to a staggered pattern of fermionic density between the sites from complementary sublattices~\cite{semenoff-1984:PRL}. Then a lattice-regularized Hamiltonian for HODF, following the spirit of Eq.~\eqref{eq:HOHamiltonian}, is given by
\begin{eqnarray}~\label{eq:HODFLattice}
H^{\rm latt}_{\rm HO}(\vec{q}) &=& H^{\rm latt}_{\rm Dir}(\vec{q}) + i \alpha \; M_{\rm latt} \; H^{\rm latt}_{\rm Dir}(\vec{q})\nonumber \\
&\equiv& t \left( 
\begin{array}{cc}
0 & (1+i \alpha)f(\vec{q}) \\
(1-i \alpha)f^\star(\vec{q}) & 0
\end{array}
\right),
\end{eqnarray}
which on the graphene lattice corresponds to a NN hopping $t_{AB}= \tilde{t} \exp(i \phi)=t^\star_{BA}$, with the amplitude ${\tilde t}=t \sqrt{1+\alpha^2}$ and the phase factor $\phi=\tan^{-1}(\alpha)$.

Expanding $H^{\rm latt}_{\rm Dir}(\vec{q})$ around two inequivalent corners of the Brillouin zone, located at $\pm {\bf K}$ (also known as the valleys), where ${\bf K}=(1/2, 1/2\sqrt{3}) 4\pi/\sqrt{3}a$, one immediately arrives at the continuum description of the massless HODF, shown in Eq.~\eqref{eq:HOHamiltonian}, with $v_{_{\rm F}}=\sqrt{3}t a /2$, $\Gamma_1=\tau_3 \beta_1$, $\Gamma_2=\tau_0 \beta_2$, and $M=\tau_0 \beta_3$, where the newly introduced set of Pauli matrices $\{ \tau_\mu\}$ act on the valley indices. It should be noted that graphene fosters a plethora of mass orders~\cite{Ryu-Chamon-Hou-Mudry-2009:PRB, Szabo-Roy-2021:PRB} and the matrix operator corresponding to any such order can be chosen as $M$ in Eq.~\eqref{eq:HOHamiltonian} and $M_{\rm latt}$ in Eq.~\eqref{eq:HODFLattice} to construct the Hamiltonian for HODF.

\emph{Summary and discussions}.~A Lorentz symmetric Hermitian extension of the Dirac Hamiltonian with an accompanying \emph{masslike} Dirac operator that also vanishes as momentum $\vec{k} \to 0$ is proposed, in which the gapless quasiparticle excitations possess an enhanced Fermi velocity ($v^{\rm R}_{_{\rm F}}$). Thermodynamic (DoS and specific heat), transport (optical conductivity), and elastic (optical shear viscosity) responses of a collection of such \emph{symmetry-protected} unconventional gapless fermionic excitations are \emph{identical} to those for conventional Dirac fermions, however, in terms of $v^{\rm R}_{_{\rm F}}$. Hidden ordering is shown to defer the onset of any symmetry breaking mass ordering, the degree of which depends on the internal algebra between the associated order parameter matrix and the masslike Dirac Hamiltonian. While a Lorentz symmetry emerges generically in the vicinity of any mass ordering, the associated GNY-QCP either continues to display hidden ordering (CCM ordering) or it disappears from the system (ACM ordering). In both cases, I extract the critical exponents at the GNY-QCP that is devoid of any sharp quasiparticle excitations and supports a non-Fermi liquid (marginal Fermi liquid) in $d=2$ ($d=3$).

Honeycomb lattice-based concrete models for HODF allows the predictions to be tested from exact diagonalization, self-consistent Hartree-Fock mean-field calculations, and quantum Monte Carlo simulations without encountering the infamous sign problem with on-site Hubbard~\cite{QMC:1, QMC:2, QMC:3} and nearest-neighbor Coulomb repulsions~\cite{QMC:4, QMC:5}, which has been studied extensively in pristine graphene and recently in the context of non-Hermitian Dirac fermions, realized by taking $\alpha \to i \alpha$ in Eqs.~\eqref{eq:HOHamiltonian} and~\eqref{eq:HODFLattice}~\cite{NH:1, NH:2, NH:3, NH:4}. The principle of constructing a hidden-ordered Hamiltonian is not limited to Dirac systems, rather it can be introduced in any lattice model in any dimension that supports gapless fermionic excitations and on which $M_{\rm latt}$ can be identified. Such a possibility opens a fascinating future direction of scrutinizing the impact of hidden order on spontaneous symmetry breaking (such as the inverse catalysis), where the DoS is constant (Fermi liquids) or diverging (flat bands).

Simplicity of the proposed lattice model for HODF, in which the hidden ordering stems from purely \emph{imaginary} reciprocal NN hopping [see Eq.~\eqref{eq:HODFLattice}], make them realizable on the highly tunable platform of optical honeycomb lattices on which purely imaginary next-NN hopping has already been emulated~\cite{opticalgraphene:1} and the strength of the Hubbard repulsion can be tuned desirably~\cite{opticalgraphene:2}. On optical lattices, various other crystalline setups can be realized (including a cubic lattice)~\cite{opticalgraphene:3, opticalgraphene:4}; altogether constituting a laboratory-based setup where impact of hidden ordering on various spontaneous symmetry breaking with special focus on their \emph{inverse catalysis} can be tested. Furthermore, recent progress in engineering \emph{designer graphene}~\cite{designergraphene:1, designergraphene:2, designergraphene:3}, on which the band width and interaction strength can tuned to large extent, constitutes yet another promising platform to experimentally showcase emergent phenomena of correlated HODF.

\emph{Acknowledgments}.~This work was supported by NSF CAREER Grant No.\ DMR-2238679 of B.R. I thank Vladimir Juri\v{c}i\'c and Christopher A.\ Leong for critical reading of the manuscript, and Igor F.\ Herbut for continued interest in this work. 

\emph{Data availability}.~No data were generated during the course of this project.

\end{document}